\documentclass[preprint2]{aastex}

\shorttitle{The hard X-ray spectrum of NGC~1365}
\shortauthors{L.~Miller \& T.J.~Turner}

\begin{document}

\title{The hard X-ray spectrum of NGC~1365: scattered light, not black hole spin}

\author{L.~Miller}
\affil{Dept. of Physics, Oxford University, 
Denys Wilkinson Building, Keble Road, Oxford OX1 3RH, U.K.}

\and

\author{T.~J.~Turner}
\affil{Dept. of Physics, University of Maryland Baltimore County, Baltimore, MD 21250, U.S.A.}

\begin{abstract}
Active Galactic Nuclei (AGN) show excess X-ray emission above 10\,keV
compared with extrapolation of spectra from lower energies.
\citeauthor{risaliti13a} have recently attempted to model the hard
X-ray excess in the type~1.8 AGN NGC~1365, concluding that the hard
excess most likely arises from Compton-scattered reflection of X-rays
from an inner accretion disk close to the black hole. Their analysis
disfavored a model in which the hard excess arises from a high column
density of circumnuclear gas partially covering a primary X-ray
source, despite such components being required in the NGC~1365 data
below 10\,keV.  Using a Monte Carlo radiative transfer approach, we
demonstrate that this conclusion is invalidated by (i) use of slab
absorption models, which have unrealistic transmission spectra for
partial covering gas, (ii) neglect of the effect of Compton scattering
on transmitted spectra and (iii) inadequate modeling of the 
spectrum of scattered X-rays. The scattered spectrum is geometry dependent
and, for high global covering factors, may dominate above 10\,keV.  We
further show that, in models of circumnuclear gas, the
suppression of the observed hard X-ray flux by reprocessing may be no
larger than required by the `light bending' model invoked for inner
disk reflection, and the expected emission line strengths lie within
the observed range.  We conclude that the time-invariant `red wing' in
AGN X-ray spectra is probably caused by continuum transmitted through and 
scattered from circumnuclear
gas, not by highly redshifted line emission, and that measurement 
of black hole spin is not possible.
\end{abstract}


\section{Introduction}

Observations of AGN reveal evidence for absorbing gas surrounding
an accreting black hole system at both X-ray \citep[see][for a review]{turner09c} and
UV wavelengths \citep[e.g.][]{crenshaw03a}.
High column density outflowing gas may be seen in up to 40\% of type\,1 AGN
\citep[e.g.][]{cappi09a, tombesi12a}.
A recent {\it Suzaku}  study \citep{tatum13a} of a hard X-ray selected sample of type~1 
AGN has shown luminosities above 10\,keV to be much higher than expected from extrapolation 
of models from the 0.5--10\,keV band.  
This `hard excess' is ubiquitous in local type~1 AGN and, owing to the sharp absorption edges 
associated with the hardest source spectra, cannot
be explained other than by the presence of a Compton-thick layer of
low-ionization absorbing gas, covering a large fraction of the
continuum source \citep{tatum13a}. 

Partial-covering absorption by gas below the
Compton-thick limit has long been known in type~1 AGN
\citep{piro05a, miller08a} and changes in the 
absorber may explain X-ray spectral
variability on long timescales. The \citet{tatum13a} analysis 
extends the paradigm into the Compton-thick regime.

It has also long been claimed that X-ray spectra 
show features caused by Compton scattering and
absorption (known as `reflection') from
an inner accretion disk, blurred by relativistic effects.
Spectral curvature 
over the 2--8\,keV band may be fit with a model that  
convolves reflection spectra \citep{ross05a} with
general relativistic effects \citep[c.f.][]{laor91a}.

However, it is difficult to distinguish absorption and reflection signatures by fitting 
individual X-ray spectra,
because Compton scattering and absorption shape both models, 
leading to similarities in overall shape. 
Consideration of AGN variability can help, and
perhaps the biggest problem for the blurred reflection-dominated
picture is the lack of any clear correlation between
continuum and reflection flux in variable AGN (section\,\ref{sec:discussion}).

A particularly interesting case is
the type~1.8 AGN \object{NGC~1365}, which has been observed extensively at 0.5--10\,keV
\citep{risaliti00a, risaliti05b, risaliti05a, risaliti07a, risaliti09c, risaliti09b}
and higher energies \citep{risaliti09a, risaliti13a}.
These observations have shown spectral variability on timescales as short as a few hours,
attributed to changes in the X-ray absorber.

\citet{risaliti13a} present simultaneous {\it XMM-Newton} and {\it NuSTAR} data for \object{NGC~1365}, 
covering 0.5--79\,keV. The authors claim that the data show relativistic
disk features: broad Fe K$\alpha$ line emission and
Compton scattering excess above 10\,keV. \citeauthor{risaliti13a} claim that the reflected component  
arises within 2.5 gravitational radii of a rapidly
spinning black hole and that absorption-dominated models that do not
include relativistic disk reflection can be ruled out. 
As this is a very strong assertion, we investigate here the assumptions 
upon which it is based. 

There are two key processes that shape AGN X-ray spectra: 
Compton scattering and photoelectric absorption.  
At the high energies observed by {\it NuSTAR}, and for 
Compton-thick gas, it is crucial to include both effects.
\citet{risaliti13a} created models representing slabs of 
ionized absorbing gas. 
However, those models fitted to \object{NGC~1365} required column density 
$N_{\rm H} \sim 5 \times 10^{24} {\rm cm^{-2}}$. Such gas would be Compton-thick,
but current ionized absorption models do not include Compton scattering. 
The circumnuclear gas also results in production of a scattered X-ray component which
is highly dependent on the geometry of the scattering region.

In this paper we present Monte Carlo simulations that demonstrate the importance of the inclusion of Compton 
scattering for the column density regime applicable to \object{NGC~1365}. 
We also demonstrate the strong geometry-dependence of the predictions for both transmitted and scattered X-rays 
and discuss the implications for modeling and interpretation of AGN X-ray spectra.

\section{Absorption and scattering by a spherical cloud}\label{sec:sphericalcloud}

\subsection{The importance of geometry}
In this section, we consider a single cloud
of constant density, absorbing photons from a distant source,
to demonstrate the key effects that must be taken into account.
The gas is assumed to be cold, with solar abundances \citep{anders89a} 
and cross-sections of \citet{verner96a}.
We first consider the effect of cloud geometry alone, omitting 
the effects of Compton scattering, for a cloud with column density $5\times 10^{24}$\,cm$^{-2}$,
as found by \citet{risaliti13a}. The cloud is illuminated by
a power-law of photon index $\Gamma=2$ and has a normalization, that would be seen by the
observer in the absence of absorption, of unity at 1\,keV.
Fig.\,\ref{fig:sphericalcloud} plots the transmitted spectrum $E f(E)$, 
for photon energy $E$ and energy spectral flux density $f(E)$.
The lowest dashed curve shows
the standard calculation obtained for a plane slab of uniform density, where 
the transmitted flux $f(E) \propto \exp(-N_{\rm H}\sigma(E))$ for 
photoelectric absorption cross-section per hydrogen atom $\sigma(E)$. 

However, a plane slab is an unlikely choice for a clumpy distribution of gas that partially
covers the source. The distribution of column density is likely
to be complex, and it is instructive to compare with a uniform spherical cloud.  
Now, there is a distribution of column density associated
with the cloud, $p(N_{\rm H}) = 2N_{\rm H}/N_{\rm H, max}^2$ for $N_{\rm H} \le N_{\rm H,max}$, 
where $N_{\rm H,max}$ is the maximum column density through the cloud.  
Integrating over this distribution, and in the absence of Compton scattering, 
we expect transmitted flux 
\begin{equation}
f(E) \propto \frac{
2\left(1-\left[1+N_{\rm H, max}\sigma(E)\right]{\rm e}^{-N_{\rm H, max}\sigma(E)}\right)}
{\left(N_{\rm H, max}\sigma(E)\right)^2}\nonumber.
\end{equation}
The upper dashed line in Fig.\,\ref{fig:sphericalcloud} shows
the results for a cloud with mean density the same as the slab 
and for which $N_{\rm H, max}$ is 3/2 times the mean column density.  At high optical depths the
transmitted flux $f(E) \rightarrow 2\left(N_{\rm H, max}\sigma(E)\right)^{-2}$, and since $\sigma(E)$
has an approximately power-law form, a hard, powerlaw spectrum develops.  In this regime, the
dominant contribution to the transmitted flux comes from the periphery of the cloud.

The spherical cloud has a significantly different transmitted spectrum from the slab. 
Such a cloud model has the same number of model parameters as the slab, and is
likely to be a better representation of a partial-covering cloud.  
This calculation alone demonstrates how careful one must be when fitting
absorption models to AGN X-ray spectra.

\subsection{The effect of Compton scattering}
At high column densities, incident radiation is Compton scattered out of the line of sight.  
To model the transmitted and scattered spectra, we carried out a Monte Carlo 
radiative transfer simulation, following \citet{miller09a}.  
Multiple scatterings were allowed, and 
when photons were absorbed by a K shell of Fe, an Fe K line photon could be produced.
Angle-dependent Klein-Nishina Compton scattering was calculated.
The Monte Carlo spectra have been smoothed with 
a gaussian of FWHM $\Delta E/E = 0.0283$, appropriate for the resolution of
typical CCD spectral data around Fe\,K. 

\begin{figure}
\begin{center}
\resizebox{0.5\textwidth}{!}{
\includegraphics[viewport = 50 30 600 520]{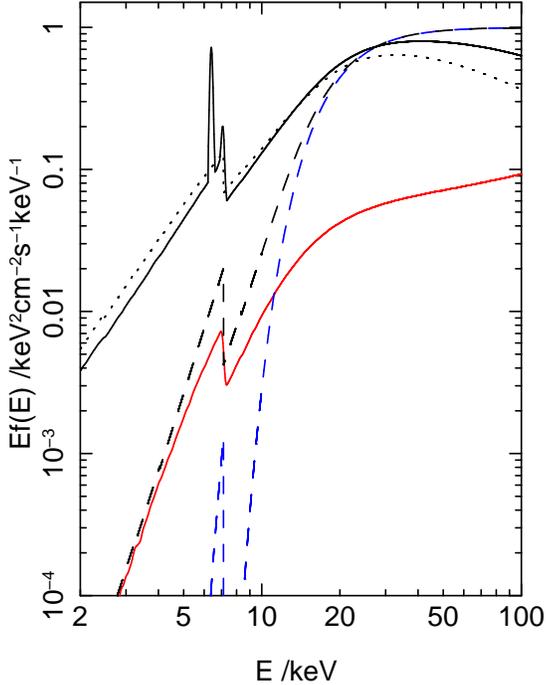}}
\end{center}
\caption{
Transmitted and scattered spectra from a single cloud:
{\em lower, dashed}: transmitted spectrum through
a uniform plane slab, neglecting Compton scattering;
{\em upper, dashed}: transmitted spectrum through a
uniform spherical cloud of same mean density as the plane slab,
neglecting Compton scattering;
{\em lower, solid}: transmitted spectrum through a uniform
spherical cloud, including Compton scattering;
{\em upper, solid}: scattered spectrum from the uniform spherical cloud;
{\em dotted}: reflection from a plane slab calculated by {\sc pexrav}.
}
\label{fig:sphericalcloud}
\end{figure}

The transmitted spectrum is shown by the lower, solid, red line in 
Fig.\,\ref{fig:sphericalcloud}.  At high energies the transmitted flux is significantly
attenuated by Compton scattering.  However, the spectrum
is significantly different from either the slab or sphere when
Compton scattering is neglected.  At low energies, the transmitted flux is again dominated
by light leaking through the periphery of the cloud, so that the spectrum
tends towards the same form as the spherical, non-scattering cloud.  

However, the light that is lost due to Compton scattering reappears in the scattered spectrum, shown
in Fig.\,\ref{fig:sphericalcloud} by the upper solid curve.  The spectrum has been integrated
over all solid angles, so that the normalization is such that this is the scattered spectrum that would be 
measured by an observer if the primary source were fully surrounded by such clouds, but neglecting
the effect of radiation being reabsorbed or scattered by other clouds (which
will be discussed in section\,\ref{sec:cloud_distribution}).  If instead only some
fraction of lines of sight are intercepted by clouds, the normalization
should be multiplied by the global covering factor.  For any global covering factor
$C_{\rm G} \ga 0.1$, the scattered spectrum dominates over the transmitted spectrum, and both
spectra have considerably different shapes from the non-scattering slab and sphere.
It is worth noting the low equivalent width of the Fe\,K$\alpha$
emission line \citep{miller09a, yaqoob10a}.

Finally, Fig.\,\ref{fig:sphericalcloud} shows a comparison with reflection 
from a plane, neutral, optically-thick slab 
as calculated by the {\sc pexrav} model \citep{magdziarz95a} within {\sc xspec} \citep{arnaud96a},
without calculation of the emission line strength. 
The normalization and shape of the {\sc pexrav} spectrum varies with inclination: 
the curve shows the case for $\cos\theta=0.5$, which matches
most closely the cloud scattered spectrum around 10\,keV. The {\sc pexrav} spectrum
is qualitatively similar to the cloud spectrum, although there are differences in shape, 
especially above 20\,keV when the cloud spectrum has a less pronounced
`Compton hump'.  Given also the significant differences in the 
transmitted spectra, we cannot expect to model even simple Compton-thick clouds with combinations of
absorption by plane slabs and planar reflection.  We shall see in
the next section that, in models with a high covering factor of clumpy circumnuclear gas, the scattered spectrum
shows much larger departures from the {\sc pexrav} model. 
The importance of geometry for the 
observed spectrum has previously been demonstrated by \citet{murphy09a} for a smooth toroidal reprocessor.

\section{Scattering and absorption by a complex distribution of circumnuclear gas}\label{sec:cloud_distribution}
\begin{figure*}
\begin{center}
\resizebox{0.9\textwidth}{!}{
\includegraphics[clip=true, viewport= 150 250 480 550]{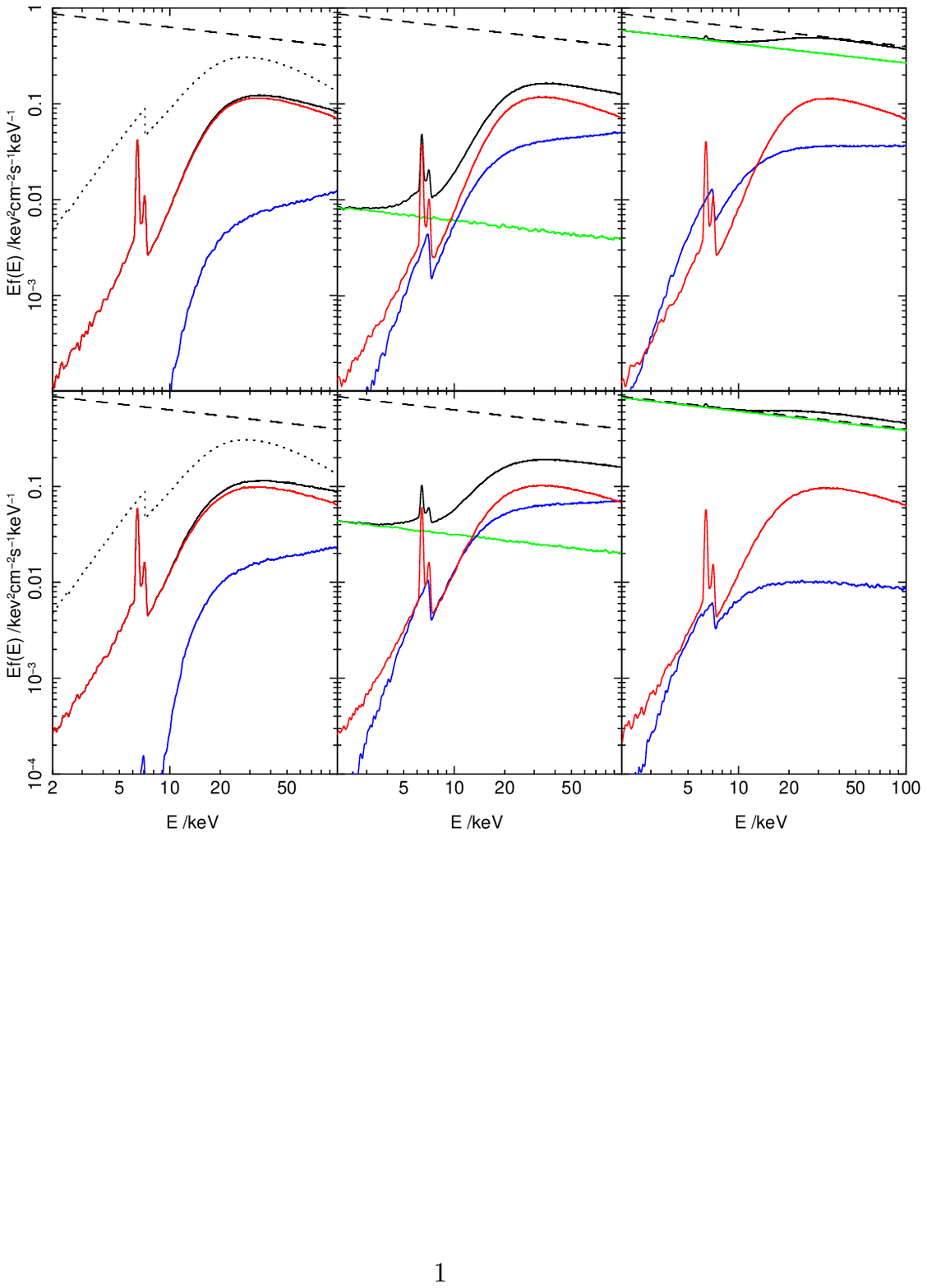}
}
\end{center}
\caption{
Spectra from the models of circumnuclear gas in section\,\ref{sec:cloud_distribution} for
models\,A (upper panels) and B (lower panels).
Each panel presents spectra from a different azimuthal angle, selected to show, from left-to-right, 
a low, intermediate and high flux state. 
The panels show the incident continuum (dashed powerlaw line), 
the spectrum of photons that have passed through 
holes in the gas distribution (green powerlaw line, not present in the left panels), 
the absorbed spectrum transmitted through the gas (blue curve),
the spectrum of scattered light, including Fe\,K line emission (red curve), 
and the total spectrum (black solid curve).  The left panels
also show a {\sc pexrav} model (dotted curve).
}
\label{fig:scatteringmodel}
\end{figure*}

In this section we consider the case of a high global covering factor of clumpy gas, 
when scattering of photons between clouds cannot be ignored (\citealt{nandra94b}, see also \citealt{merloni06a}).
As in \citet{tatum13a}, a simple, clumpy distribution was created by
placing 1000 spherical clouds of equal radius
between inner and outer radii $r_{\rm min}$ and $r_{\rm max}$ from a central X-ray source.
Gas was placed inside each cloud with a uniform density (the density was not increased within overlapping
clouds).  For illustration, we chose 
$N_{\rm H} = 1.5 \times 10^{24}$\,cm$^{-2}$ per unit cloud radius, 
chosen to approximately maximize the hard excess.
In order to simulate the effect that in type\,2 AGN we might be looking through a denser
distribution of clouds than in type\,1 AGN, the number density of clouds increased towards the equatorial
plane as $\sin\phi$, where $\phi$ is the polar angle of observation.
In Fig.\,\ref{fig:scatteringmodel} we show two models: model\,A has $r_{\min}=10$, $r_{\rm max}=20$
in units of the cloud radius; model\,B has $r_{\rm min}=10$, $r_{\rm max}=30$.
The fraction of sightlines that were empty of gas was $0.35$ at $\phi \simeq 0$, 
decreasing to $<0.15$ for $\phi>60^{\circ}$ in model\,A, and $0.55$ at $\phi \simeq 0$
decreasing to $<0.25$ for $\phi>60^{\circ}$ in model\,B.
We emphasize that these are simplified models of circumnuclear material, designed to illustrate the
radiative transfer effects, and are not to be taken as a proposal for the actual gas distribution.

Spectra were created using $6 \times 10^{10}$
input quanta with energy $0.1-400$\,keV.  The input spectrum was a powerlaw of photon index
$\Gamma = 2.2$, as inferred for \object{NGC~1365} \citep[e.g.][]{risaliti09a}.  
It was assumed that photons that propagated through the equatorial plane were absorbed
by an accretion disk and, to avoid possible confusion with disk reflection,
the disk did not re-emit photons.
As in section\,\ref{sec:sphericalcloud} the gas was assumed cold and
angle-dependent Klein-Nishina Compton scattering was calculated.  
Spectra can be accumulated only by averaging over finite solid angle, and here 
we adopted a solid angle corresponding to that subtended 
from the central source by a cloud at $r_{\rm max}$.
A point source was placed at the center of the
cloud distribution, but the averaging of photons over solid angle has the effect of mimicking the partial
covering of an extended source of comparable size to a cloud.  

Fig.\,\ref{fig:scatteringmodel} shows example spectra obtained for models\,A\,and\,B, 
viewed at polar angle $\phi = 60^{\circ}$, for three azimuthal angles.
The spectra have been smoothed as in section\,\ref{sec:sphericalcloud},
but Monte Carlo noise is visible at low flux values.
The spectra cover a range similar to that observed in the AGN population, from 
`reflection dominated', through intermediate states where primary continuum is visible, to high states where
the primary continuum dominates. All states have significant hard excesses
above 20\,keV. 
Unobscured sightlines have
total light at high energy that exceeds the input continuum amplitude. The panels
show the contributions from various components: photons that pass through the 
holes in the distribution, photons that pass through clouds but are not scattered, and scattered
photons, including Fe\,K$\alpha$ and Fe\,K$\beta$ line emission.  The incident continuum 
is also shown as a dashed line. The chief results are as follows.
\begin{list}{}
  { \setlength{\itemsep}{0pt}
     \setlength{\parsep}{1pt}
     \setlength{\topsep}{1.75pt}
     \setlength{\partopsep}{0pt}
     \setlength{\leftmargin}{0.em}
     \setlength{\labelwidth}{0.em}
     \setlength{\labelsep}{0.2em} }
\item[(i)] The transmitted, unscattered spectrum (blue curves) is variable 
and does not look like the spectrum expected from transmission through a 
slab model (section\,\ref{sec:sphericalcloud}).
\item[(ii)] The scattered radiation is different from that calculated 
in either the isolated cloud or {\sc pexrav} models (the left panels of
Fig.\,\ref{fig:scatteringmodel} shows a {\sc pexrav} model for photon index $\Gamma=2.2$ and
$\cos\theta=0.5$).  The hardness of the scattered spectrum increases with global covering factor.
\item[(iii)] Within each model, the scattered radiation is invariant with 
azimuthal viewing angle, because it does not depend on
whether the primary source happens to be covered, or not, along any one sightline.
\item[(iv)] In all but the highest flux states, 
the scattered light is the dominant spectral component above 20\,keV,
where the total amplitude is only a factor 3--4 below the incident continuum, even
in this regime where the $2-10$\,keV continuum may be suppressed by much larger factors.
\item[(v)] The line flux of Fe\,K$\alpha$ is low, as expected \citep{miller09a, yaqoob10a}.  
Equivalent widths vary from 1500\,eV in the reflection-dominated state down to 10\,eV in the
highest state. The equivalent width of Fe\,K$\alpha$ found in the three 
Suzaku observations of \object{NGC~1365} is 330--520\,eV (Tatum et al., in preparation).
\end{list}

\section{Discussion}\label{sec:discussion}

\subsection{The `red wing' in AGN spectra}
The results of section\,\ref{sec:cloud_distribution} show that, on long timescales where spectral variability
may be dominated by variations in observer covering fraction, there should be a spectrum of largely invariant
scattered light that creates an apparent `red wing' at energies below the Fe\,K edge.  Such an invariant
red wing is well known \citep[e.g.][]{fabian03a}, and
using principal components analysis, \citet{miller07a, miller08a, miller10a}
showed that there is an approximately invariant `offset' component in AGN spectra. 
The effect leads to a problem with the
explanation of the red wing as reflection by the inner accretion disk.  If the
disk is close to the primary X-ray source, its flux variations should be
closely tracked by the reflected light, but apparently they are not. \citet{fabian03a}, \citet{miniutti03a}
and subsequent papers proposed a `light bending' model in which a compact, primary source near the black
hole moves towards or away from the disk, such that bending of light causes 
an apparent variation in flux from the primary source, but not from the reflected light.
This model has to be finely tuned to obtain
invariant reflected flux \citep{miniutti04a}.

Such an {\em a posteriori} model is not required if 
the red wing arises from a combination of transmitted and scattered continua, 
as suggested by \citet{turner07a}, and not from highly redshifted line emission.
\citet{merloni06a} have also suggested this
explanation in the context of scattered light arising in a model of an inhomogeneous accretion flow.

\subsection{Modeling the spectrum of NGC~1365}
In modeling the spectrum of \object{NGC~1365}, \citet{risaliti13a} 
recognize that partial covering absorption
is required, regardless of whether there is also a component of blurred reflection.
In an initial comparison (Fig.~3 of \citealt{risaliti13a}), 
a model of blurred reflection plus one absorber component
was compared against a model of two absorber components.  The models were only fit to the {\em XMM-Newton}
data below 10\,keV, and it was then seen how well each model extrapolated to higher energy.
However, the blurred reflection included Compton
scattering, but the two-absorber model did not.  Hence it is no surprise that the first
model extrapolated better above 10\,keV.  In a subsequent test,
\citet{risaliti13a} fit also to the hard X-ray {\em NuSTAR} data and claimed that the blurred model is
statistically better. But the transmitted and scattered light components were modeled
with non-scattering, slab absorbers plus a {\sc pexrav} component, so such a test is invalid.
Point (ii) above demonstrates that existing models such as {\sc pexrav}
cannot accurately account for reprocessing by a complex circumnuclear gas distribution.

Estimates of the continuum suppression arising from Compton scattering require proper inclusion
of light scattered back into the line of sight by circumnuclear gas.
\citet{risaliti13a} considered the expected effect for Compton optical depth $\tau \simeq 3-4$, using the {\sc plcabs} model \citep{yaqoob97a}, and inferred that the intrinsic continuum would be a factor 8 larger than the observed continuum.  The problems with this approach are:\\
(i) the assumed Compton depth was obtained from fitting slab absorption models, 
which is likely to be incorrect and thus lead to a significant error in the Compton loss estimate;\\
(ii) the approximations used in {\sc plcabs} are not valid above 10\,keV for high column density 
\citep{yaqoob97a};\\
(iii) the {\sc plcabs} model assumes full covering by a spherical shell, 
and we do not expect this to yield the same scattered light spectrum as clumpy, partial 
covering gas - our Monte Carlo simulations supersede the use of {\sc plcabs} 
in this context.\\ 
In our models, the Compton loss correction is only a factor 3--4 
at high energy, comparable to that
in the `light bending' model of inner disk reflection, 
where up to 75\% of light is lost \citep{miniutti04a}.

The scattering models presented here also show that the expected equivalent width of Fe\,K$\alpha$ covers
a wide range that includes the observed range for NGC~1365.
Finally, \citet{tatum13a} argue that all type\,1 AGN are affected by high-column, partial covering
gas distributions, so 
the ratio of X-ray to infrared or optical line emission for NGC 1365 would not be expected to appear
anomalous.

\subsection{On the difficulty of measuring black hole spin}
It has been claimed that black hole spin can be measured from the shape of the ``red wing'',
whose profile varies as a function of spin parameter $a$, which
requires red wing measurement down to energies
$\la 3$\,keV, free from the effects of other modifications to the continuum.
Even in the absence of circumnuclear reprocessing, $a$ is difficult to measure free
from degeneracies with other parameters \citep{dovciak04b}.
Given that in AGN such as NGC~1365, Mrk~766 
\citep{miller07a} and MCG--6-30-15 \citep{miller08a} the entire red wing may be modeled as
scattered and absorbed continuum, and given that complex absorption must
be present \citep[e.g. NGC~1365,][]{risaliti13a}, it is not possible to
unambiguously identify and measure 
any alleged blurred red wing component with sufficient accuracy to
determine spin in any of the AGN that have been subject to detailed study to date.

\section{Conclusions}
AGN spectra may contain significant levels of transmitted and scattered continua from
partially-covering circumnuclear gas, producing the time-invariant `red wing'
below the Fe\,K edge and modifying the spectrum above
10\,keV.  Such gas is already known in NGC~1365.
Geometry-dependent Compton scattering must be taken into
account, using 3D radiative transfer calculations, when making spectral models and when
accounting for the source total luminosity.  The expected red wing, low Fe\,K$\alpha$ equivalent width and high
energy excess may be easily confused with `reflection' from
an accretion disk, and spectral analyses to date have not established
that there is any requirement for inner-disk reflection. Improved models of clumpy, ionized circumnuclear
gas are required to achieve a full understanding of AGN X-ray spectra.

\acknowledgments
TJT acknowledges NASA grant NNX11AJ57G.


\end{document}